\documentclass[prl,twocolumn,showpacs,amsmath,amssymb,superscriptaddress,floatfix]{revtex4}

\usepackage{graphicx}% include figure files
\usepackage{bm}% bold math

\begin{document}

%\preprint{Preprint \ldots}

\title{Broken Time Reversal of Light Interaction with Planar Chiral Nanostructures}

\author{A. S. Schwanecke}
\affiliation{Department of Physics and Astronomy, University of Southampton, SO17 1BJ, UK}
\author{A. Krasavin}
\affiliation{Department of Physics and Astronomy, University of Southampton, SO17 1BJ, UK}
\author{D. M. Bagnall}
\affiliation{Department of Physics and Astronomy, University of Southampton, SO17 1BJ, UK}
\affiliation{Department of Electronics and Computer Science, University of Southampton, SO17 1BJ, UK}
\author{A. Potts}
\affiliation{Department of Physics and Astronomy, University of Southampton, SO17 1BJ, UK}
\author{A. V. Zayats}
\affiliation{School of Mathematics and Physics, Queen's University of Belfast, BT7 1NN, UK}
\author{N. I. Zheludev}
\email[Email: ]{n.i.zheludev@soton.ac.uk} \affiliation{Department of Physics and Astronomy, University of Southampton, SO17 1BJ, UK}
\date{\today}

\begin{abstract}
We report unambiguous experimental evidence of broken time reversal symmetry for the interaction of light with an artificial non-magnetic material. Polarized colour images of planar chiral gold-on-silicon nanostructures consisting of arrays of gammadions show intriguing and unusual symmetry: structures, which are geometrically mirror images, loose their mirror symmetry in polarized light. The symmetry of images can only be described in
terms of anti-symmetry (black-and-white symmetry) appropriate to a time-odd process. The effect results from a transverse chiral non-local electromagnetic response of the structure and has some striking resemblance with the expected features of light scattering on anyon matter.
\end{abstract}

%%%PACS notes
%%%10. THE PHYSICS OF ELEMENTARY PARTICLES AND FIELDS (for cosmic rays, see 96.40.-z in astronomy; for experimental methods and instrumentation for elementary-particle physics, see section 29)
%%%11. General theory of fields and particles (see also 03.65.-w Quantum mechanics and 03.70.+k Theory of quantized fields)
%% ==> 11.30.-j Symmetry and conservation laws (see also 02.20.-a Group theory)
%%%40. ELECTROMAGNETISM, OPTICS, ACOUSTICS, HEAT TRANSFER, CLASSICAL MECHANICS, AND FLUID DYNAMICS
%%%42. Optics (for optical properties of gases, see 51.70.+f; for optical properties of bulk materials and thin films, see 78.20.-e; for x-ray optics, see 41.50.+h)
%%%42.25.-p Wave optics
%% ==> 42.25.Ja Polarization
%%%70. CONDENSED MATTER: ELECTRONIC STRUCTURE, ELECTRICAL, MAGNETIC, AND OPTICAL PROPERTIES
%%%71.  Electronic structure of bulk materials (see section 73 for electronic structure of surfaces, interfaces, low-dimensional structures, and nanomaterials; for electronic structure of superconductors, see 74.25.Jb)
%% ==> 71.10.Pm Fermions in reduced dimensions (anyons, composite fermions, Luttinger liquid, etc.) (for anyon mechanism in superconductors, see 74.20.Mn)
%%%78. Optical properties, condensed-matter spectroscopy and other interactions of radiation and particles with condensed matter
%% ==> 78.67.-n Optical properties of low-dimensional, mesoscopic, and nanoscale materials and structures



\maketitle

Light-matter interactions involving non-magnetic materials are generally believed to obey time reversal symmetry, which is seen as a direct consequence of the time-invariance, when taken separately, of non-magnetic media and the Maxwell equations. However, when the reversability argument is considered, the respective orientations of the electromagnetic wave propagation direction and the medium it interacts with may also be important, for instance in the case of planar chiral structures (PCS). A planar chiral structure is a pattern that cannot be brought into congruence with its mirror image unless it is lifted from the plane. In essence a planar chiral structure has a perceived sense of `twist'. In 1994, Hecht and Barron, noted that the sign of the twist reverses if the structure is observed from different sides of the plane, and that this should also be the case for light polarization effects associated with the structure \cite{hecht1994}. The possibility that this could imply the breaking of time reversal symmetry has been discussed in our recent papers \cite{papakostas2003,prosvirnin2003}. Here, we report on experimental evidence of a time non-reversal interaction between light and an artificial non-magnetic material, namely metallic planar chiral nanostructures.

The planar chiral structures belong to a new type of optical meta-material that have not until recently been systematically investigated \cite{papakostas2003}. PCS studied here consist of regular arrays of chiral gammadions with a groove width of about 700nm cut into a thin film of metal (100nm of gold sandwiched between two 20nm thick layers of titanium) and arranged to produce a pattern of planar 442 wallpaper group symmetry. More details on the structures may be found in \cite{papakostas2003} where we reported that they affect the polarization state of diffracted light in an enantiomerically-sensitive fashion. The failure of time reversal symmetry for optical interactions with the planar chiral structures, which we report here, are evidenced by the unusual symmetries observed in images of the structures obtained using a polarizing optical microscope. The observations were performed in reflective mode, with a white light halogen source, using a $40\times$ microscope objective and a 6.3 megapixel low noise CMOS CCD camera. Light incident on the structure was linearly polarized and the sample was imaged through a `crossed' linear analyser, giving a dark field for unstructured metallic surfaces.

\begin{figure}[!t]
\includegraphics[scale=0.50]{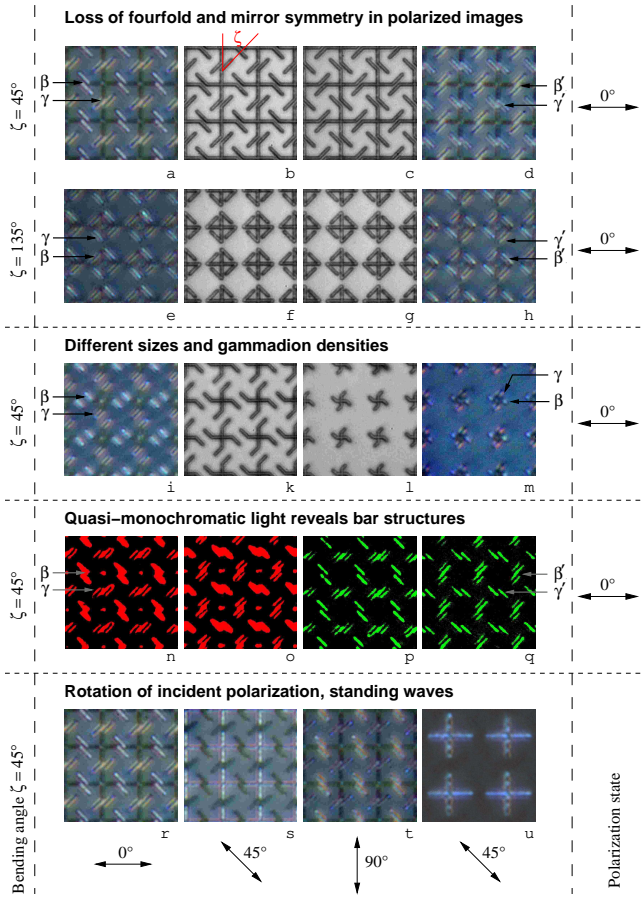}
\caption{\label{fig-microscope}[colours online] Intriguing colour symmetry of images of metallic planar chiral structures taken in polarized light (\texttt{a}, \texttt{d}, \texttt{e}, \texttt{h}, \texttt{i}, \texttt{m} -- \texttt{u}). Grayscale images \texttt{b}, \texttt{c}, \texttt{f}, \texttt{g}, \texttt{k}, \texttt{l} were taken in unpolarized light to show the underlying topography of the adjacent colour images. All images are $30\mu{}m\times{}30\mu{}m$ apart from image \texttt{u}, which is $20\mu{}m\times20\mu{}m$.}
\end{figure}

Polarized images of planar chiral metallic nanostructures are presented in Fig.~\ref{fig-microscope}. The polarization state of incident light is indicated by double-ended arrows alongside the rows of images. Images in the first row show of right- and left-handed enantiomeric (mirror) forms of the same PCS. Images \texttt{b} and \texttt{c} were taken in unpolarized light and are presented here to illustrate the topography of the enantiomeric forms. They show that geometrically the structures \texttt{b} and \texttt{c} are mirror images (enantiomeric forms) of one another and that they have fourfold rotational symmetry. Image \texttt{a} is a polarized image of structure \texttt{b}, while image \texttt{d} is a polarized image of structure \texttt{c}. In images \texttt{a} and \texttt{d} the elements of the gammadion structures oriented at 45$^\circ$ to the incident polarization direction are coloured. They are labelled $\beta$, $\gamma$ on image \texttt{a} and $\beta^\prime$, $\gamma^\prime$ on image \texttt{d}. Intriguingly (see image \texttt{a}), the 45$^\circ$--elements of the horizontal ($\beta$) and vertical ($\gamma$) branches are seen in different colours, and therefore the polarized colour images lose the fourfold rotational symmetry of the underlying structure. Moreover, the colouring of the branches is different for the two enantiomeric forms: elements of type $\beta$ are seen as single bright fields of bluish colour, while the same elements of type $\beta^\prime$ belonging to the opposite enantiomeric form are seen as dark fields surrounded by a bright yellowish border. This lack of mirror symmetry is emphasized on the contrast enhanced polarized images \texttt{n}/\texttt{o} and \texttt{p}/\texttt{q} taken in quasi-monochromatic light with spectral filters at 600nm and 530nm respectively. They show distinct double and triple bar structures at different gammadion branches. The loss of the fourfold rotation axis and the absence of any mirror symmetry in the polarized images of two enantiomeric forms is observed for gammadions of various shapes (for instance \texttt{e}/\texttt{h} are polarized light images of `closed' gammadions with a bending angle $\zeta = 135^\circ$, with topography presented in images \texttt{f}/\texttt{g}), and on structures with gammadion features of different size, as illustrated in pictures \texttt{i}/\texttt{m} (unpolarized images \texttt{k} and \texttt{l} show topography). A progressive change in the colouring of the gammadion structural elements is seen when the polarization state of the incident light is changed (images \texttt{r}, \texttt{s}, \texttt{t}). Comparison of images \texttt{r} and \texttt{t} shows that the colouring of the gammadion branches depends on the incident polarization state, and therefore is not related to any topographical manufacturing imperfection.

\begin{figure}[!t]
\includegraphics[scale=0.5]{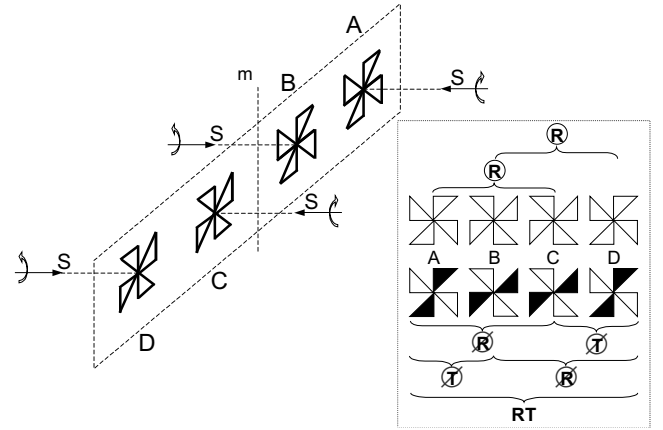}
\caption{\label{fig-symmetry}An electromagnetic wave interacting with a generic planar chiral object (scenario A), and symmetry transformations: (B) Time reversal, $\bm{T}$; (C) Enantiomeric transformation of the sample, $\bm{R}$; (D) Enantiomeric time reversal, $\bm{RT}$. Boxed part: The upper row shows the topography of the of the planar chiral object, as it appears when observed from the direction of light propagation, while the lower row shows simplified colour patterns for the corresponding polarized images.}
\end{figure}

In order to analyse the importance of these observations, one must consider the results of different symmetry transformations on the light-PCS interaction process (Fig.~\ref{fig-symmetry}). Since a plane wave may be represented as a linear combination of two circularly polarized waves, Fig.~\ref{fig-symmetry} shows a wave of one handedness for simplicity. In a generic scenario A a clockwise PCS interacts with an anti-clockwise polarized wave (looking towards the light source). Time reversal (operator $\bm{T}$) leaves the gammadion structure and handedness of the wave unchanged, but reverses the direction of wave propagation $\bm{S}$, leading to scenario B. Is the light-matter interaction time reversed? As the relative sense of rotation between the wave and the planar chiral structure is different in cases A and B, in a sub-wavelength planar chiral structure this could manifest itself as a difference in reflection, transmission and scattering. If it does, this would constitute broken reversability. In this case certain areas of the structure that we image may have different appearances (colours, brightness, polarization state) when observed from opposite sides. Let us now consider an experiment with the enantiomeric form of the planar structure as achieved by reflection about line \textit{m}, leading to scenario C (we will denote this enantiomeric conversion of the structure as operator $\bm{R}$). Again, as a result of the difference in the relative sense of rotation between the wave and the planar chiral structure there is no reason to expect that the interaction of light with the structures would give the same result in scenarios A and C, thus permitting a difference in reflection, transmission and scattering for light incident on the two enantiomeric forms of a sub-wavelength structure. When a structure is imaged, the mirror-symmetric areas of gammadions in scenarios A and C may have different appearances. However, if we now examine the time reversal scenario with the enantiomeric form of the gammadion (scenario D), the relative handedness of the wave and the planar chiral structure are identical to case A. The interaction of light with the gammadion structure will therefore give exactly the same results. Such a symmetry transformation ($\bm{RT}$) is known as enantiomeric time-reversal and is being seen as the most universal symmetry, ruling chemical kinetics \cite{barron1994} and optical interactions \cite{svirko1995} involving chiral molecules.

Therefore, the interaction of light with a planar chiral structure must obey enantiomeric time-reversal, but may or may not obey simple time-reversal symmetry. So, the question is whether we can draw conclusions on time reversal symmetry from the observations described above? And the answer is yes, indeed we can. The optical images of the generic structure in polarized light (scenario A), presented in Figures \ref{fig-microscope} \texttt{a}, \texttt{e}, \texttt{n} and \texttt{p}, and the corresponding images of its enantiomeric form (scenario C), presented in Figures \ref{fig-microscope} \texttt{d}, \texttt{h}, \texttt{o} and \texttt{q}, are distinctively different, the mirror-symmetric areas of gammadions have different appearances and even different colours. The $\bm{R}$-symmetry is, therefore, broken. Thus, the fact that the system obeys $\bm{RT}$-symmetry overall, means that $\bm{T}$-symmetry must also be broken. If the system were imaged in polarized light in the time reversed scenario, one would observe an image transformation similar to that achieved by enantiomeric transformation of the sample. In this case the most distinctive feature would be the swapping of the colours of the adjacent branches of the gammadion, as schematically illustrated in Fig.~\ref{fig-symmetry}.

The symmetry of polarized images may be described in terms of magnetic symmetry, which is closely related to black-and-white symmetry and the Shubnikov anti-symmetry which is used to describe time odd effects \cite{landau1984,shubnikov1964}. In these terms the geometrical gammadion structure has a fourfold rotation axis, while the polarized images (and therefore the light-matter interaction) has a fourfold axis of anti-rotation, meaning that to transform the image to a state indistinguishable from the original, a geometrical rotation of 90 degrees must be followed or preceded by a `repainting' of the adjacent gammadion branches (here `repainting' shall be understood as changing the branch colours from bluish to yellow and vice versa which is accompanied by switching from double to triple bar structure).

We believe that any sort of aperture effect cannot  explain the peculiar symmetrical properties and colouring of the polarized images: the same features are seen in the images irrespective of microscope objective magnification, i.e.~of the acceptance angle of the lens ($40\times$, $10\times$, $4\times$). Furthermore, the appearance of these features does not depend on what part of the microscope field of view is examined, on whether the whole sample or only a fraction of it is illuminated or on which part of the sample is imaged. In addition, the polarized image symmetry is independent of pattern density (compare images \texttt{a}, \texttt{i}, \texttt{m}, Fig.~\ref{fig-microscope}) or grating pitch. The observed effect is pronounced and robust and is easily observable with a bare eye looking into the microscope. It cannot be explained by any imperfection of the polarizers: it is not destroyed by tilting the polarizer for about $\pm{}5$ degrees, and it is not sensitive to any parallel translations of the structure in respect to the field of view. The observed effects are much less pronounced in different material configurations, such as gold-based structures without top titanium layer and are undetectable in gammadion arrays manufactured in chromium films instead of gold.

\begin{figure}[!t]
\includegraphics[scale=0.43]{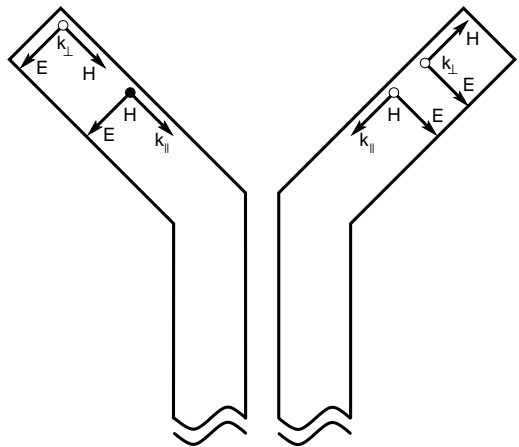}
\caption{\label{fig-calculation}Electromagnetic fields of surface plasmon-polariton waves and volume waves with wave-vectors $k_\perp$ and parallel $k_\parallel$ to the plane of the structure in two enantiomeric forms of the gammadion branch show no mirror symmetry. The `open' symbol ($\circ$) represents field vector directed out of the plane of the picture, the `closed' symbol ($\bullet$) represents field vector into the plane of the picture.}
\end{figure}

The colour patterns of polarized images arise from the excitation of interacting surface plasmon modes and standing volume waves in the grooves of the structure. To elucidate the colouring mechanism we performed 2D numerical modelling of the fields near a straight infinite groove. When a light wave of wavelength $\lambda_0$ $\left(k_0=2 \pi / \lambda_0\right)$ is incident normally on the structure, the polarization components parallel and perpendicular to the groove are split. The component polarized \textit{parallel} to the groove is \textit{not coupled} to surface plasmon excitations on the walls of the metallic structure and is reflected from the sample. The polarization component \textit{perpendicular} to the groove \textit{excites surface plasmon-polaritons} (SPP's) that have a wave vector $k_\perp$ and propagate perpendicular to the film surface on the opposing walls of the groove. Their wave-vector depends on the dielectric constant $\varepsilon$ of the metal and the groove width $d$ \cite{sobnack1998,novikov2002}. Depending on the wavelength these SPP's may be coupled through standing waves in the groove. This leads to a polarization conversion effect previously seen in narrower grooves \cite{hooper2002}: Since the wave vector $k_\perp$ is different from that in free space, the groove acts as an inhomogenous wave-plate with retardation depending upon the frequency dispersion of the dielectric characteristics of the metal film, the groove width and the position across the groove, thus giving coloured images when the structure is observed through an analyser.

The model outlined above treats the sections of a gammadion as grooves of infinite length and does not explain why different branches of the gammadions have different colours. We believe the the origin of the colour differences lies in the transverse nonlocality of the optical response, which results from the excitation of the chiral structural element as a whole, i.e.~from the coupling between different sections of gammadion branches. This collective excitation is provided by SPP's and standing volume waves excited on the walls and in the groove and propagating along the groove parallel to the plane of the structure.

The excitation of SPP's and standing volume waves with wave-vector $k_\parallel$ parallel to the plane of the structure is evident from the standing wave patterns seen on polarized image \texttt{u}, Fig.~\ref{fig-microscope}. In the case of a finite length groove the dispersion characteristics of the $k_\perp$ and $k_\parallel$ waves are mutually dependent and are both functions of the groove width and length. Importantly, coupling between $k_\perp$ and $k_\parallel$ waves is not the same in the two enantiomeric forms of the structure. This is illustrated for the case of SPP's in branches of left and right-handed gammadions in Fig.~\ref{fig-calculation}, where the momentary directions of the electric and magnetic field vectors are compared. This asymmetry results from the difference in the propagation conditions for waves travelling to and from the gammadion centre in different enantiomeric forms, and from the asymmetry of the respective orientation of the incident wave, the $\bm{E}$--$\bm{H}$--$\bm{k}$ triad and the structure's bending direction. For instance, if at a moment of time the electric fields of $k_\perp$ and $k_\parallel$ SPP's are directed towards the internal part of gammadion, and $k_\perp$ waves in both gammadions propagate upwards, the magnetic fields of the $k_\perp$ and $k_\parallel$ SPP waves propagating towards the gammadion's centre, should have anti-parallel magnetic fields in right- and left-handed branches. Therefore, there is no mirror-like field symmetry in the left- and right-handed branches, and this leads to different coupling efficiencies between the $k_\perp$ and $k_\parallel$ waves in enantiomeric structures. The coupling between the two modes makes the electromagnetic wave dispersion in the grooves functions of the sense of twist of the branch with respect to the incident light polarization. Through the wave-vector dependent retardation colouring mechanism described above this leads to the different colours of polarized images of the enantiomeric forms. The difference in colours between different branches of the same gammadion (e.g.~in image \texttt{a}, Fig.~\ref{fig-microscope}) is related to the fact that some 45$^\circ$ sections are attached to a horizontal groove in which the $k_\perp$ SPP is not excited, while other 45$^\circ$ sections are attached to a vertical groove where $k_\perp$ SPP's are efficiently excited by incident light polarized along the horizontal direction. Finally, $k_\perp$ waves in the grooves can be coupled to SPP's propagating along the structure's surface and this explains the triple bar features $\gamma$ and $\beta^\prime$. In the upper over-layer these SPP's decay rapidly because of high losses in titanium.

In conclusion, we have presented polarized images of metallic planar chiral structures, which unambiguously show evidence of time non-reversal optical interactions resulting from the transverse non-locality of the collective response of gammadion elements in the nanostructure. We point out that the main features of the observed phenomenon, namely its broken T-symmetry and two-dimensional chiral nature, have some striking phenomenological resemblance with expected peculiarities of light scattering on anyons, hypothetic quasi-particles with fractional statistics \cite{wright2003}. Essentially resulting from the two-dimensional confinement, such particals derive their name from the word \textit{any} since the phase accumulated upon exchanging two identical anyons can assume any value. They have been discussed in the theories of electrons trapped in superconducting vortices \cite{wilczek1982} and electromagnetically trapped atoms \cite{wright1994} and have been predicted to inflict time non-reversal light scattering \cite{canright1992}.

\begin{acknowledgments}
The authors acknowledge the support of the Science and Engineering Research Council (U.K.) and A. S. Schwanecke acknowledges support of the German National Merit Foundation.
\end{acknowledgments}

% The following text is directly copied from schwanecke.bbl created using BibTeX (schwanecke.bib).
%\bibliography{schwanecke}

\end{document}